\documentclass[preprint,prd,aps,showpacs,showkeys,nofootinbib]{revtex4}
\usepackage{graphicx}
\begin{document}


\title{Electron and muon $(g-2)$ in the B-LSSM}

\author{Jin-Lei Yang$^{1,2,3}$\footnote{yangjinlei@itp.ac.cn},
Tai-Fu Feng$^{1,2,4}$\footnote{fengtf@hbu.edu.cn}, Hai-Bin Zhang$^{1,2}$\footnote{hbzhang@hbu.edu.cn}}

\affiliation{Department of Physics, Hebei University, Baoding, 071002, China$^1$\\
Key Laboratory of High-precision Computation and Application of Quantum Field Theory of Hebei Province, Baoding, 071002, China$^2$\\
CAS Key Laboratory of Theoretical Physics, School of Physical Sciences,
University of Chinese Academy of Sciences, Beijing 100049, China$^3$\\
Department of Physics, Chongqing University, Chongqing 401331, China$^4$}

\begin{abstract}
The theoretical predictions in the standard model (SM) and measurements on the anomalous magnetic dipole moments (MDM) of muon and electron have great precision, hence the MDMs of muon and electron have close relation with the new physics (NP) beyond the SM. Recently, a negative $\sim2.4\sigma$ discrepancy between the measured electron MDM and the SM prediction results from a recent improved determination of the fine structure constant. Combined with the long-lasting muon MDM discrepancy which is about $\sim3.7\sigma$, it is difficult to explain both the magnitude and opposite signs of the deviations in a consistent model, without introducing large flavour-violating effects. The analysis shows that they can be explained in the minimal supersymmetric extension (MSSM) of the SM with local $B-L$ gauge symmetry (B-LSSM). Comparing with the MSSM, new parameters in the B-LSSM can affect the theoretical predictions on lepton MDMs, and the effects of them are explored.

\end{abstract}


\maketitle

\section{Introduction\label{sec1}}
\indent\indent
The anomalous magnetic dipole moments (MDM) of lepton $a_l$~\cite{Schwinger:1948iu} has been one of the most precisely measured and calculated quantities in elementary particle physics, which also provides one of the strongest tests of the SM. For the muon MDM, the discrepancy between the measured muon MDM and the SM prediction has existed for a long time~\cite{Lindner:2016bgg,Campanario:2019mjh}, which may be a hint of new physics (NP) and reads~\cite{Bennett:2006fi,Blum:2018mom}
\begin{eqnarray}
&&\bigtriangleup a_\mu\equiv a_\mu^{exp}-a_\mu^{SM}=(2.74\pm0.73)\times10^{-9}.
\end{eqnarray}
In addition, $a_\mu$ is being measured at Fermilab and J-PARC, and the upcoming results are expected to have a better accuracy.

However, a negative $\sim2.4\sigma$ discrepancy between the measured electron MDM and the SM prediction appears, due to a recent precise measurement of the fine structure constant, which changes the situation that the electron MDM is consistent with the measurement. The negative $\sim2.4\sigma$ discrepancy reads~\cite{Hanneke:2008tm,Parker:2018vye}
\begin{eqnarray}
&&\bigtriangleup a_e\equiv a_e^{exp}-a_e^{SM}=-(8.8\pm3.6)\times10^{-13}.
\end{eqnarray}

It is obvious that the signs of $\bigtriangleup a_\mu$ and $\bigtriangleup a_e$ are opposite. Even if the NP effects are considered, the MDMs of muon and electron are related without any flavor violation in the lepton sector as
\begin{eqnarray}
&&\frac{\bigtriangleup a_\mu}{\bigtriangleup a_e}\simeq m_\mu^2/m_e^2\simeq4.2\times10^4,
\end{eqnarray}
both sign and magnitude have discrepancies (which may disappear due to the latest lattice results~\cite{Crivellin:2020zul}).

In extensions of the SM, the supersymmetry is considered as one of the most plausible candidates. And the discrepancies between $\bigtriangleup a_\mu$, $\bigtriangleup a_e$ have been exhaustively studied, the results show that the discrepancies can be explained by requiring new sources of flavour violation~\cite{Giudice:2012ms,Crivellin:2018qmi,Dutta:2018fge,Calibbi:2020emz,Bigaran:2020jil}, introducing a single CP-even scalar with sub-GeV mass that couples differently to muons and electrons~\cite{Marciano:2016yhf,Davoudiasl:2018fbb,Jana:2020pxx}, introducing a light complex scalar that is charged under a global $U(1)$ under which the electron is also charged but muon not~\cite{Liu:2018xkx}, introducing axion-like particles with lepton-flavour violating couplings~\cite{Bauer:2019gfk,Cornella:2019uxs}, enhancing the SUSY electron Yukawa coupling and reverse the sign of the muon Yukawa coupling by the SUSY threshold correction in the lepton sector~\cite{Endo:2019bcj}, or requiring smuons are much heavier than selectrons to arrange the sizes of bino-slepton and chargino-sneutrino contributions differently between the electron and muon sectors~\cite{Badziak:2019gaf}. For non-supersymmetric BSM models, the authors of Ref.~\cite{Hiller:2019mou} put forward two models with new scalar and fermionic matter which can explain the discrepancies without explicit lepton flavor violation or universality violation beyond the lepton mass effects already present in the SM, and the discrepancies can also be explained in a three-loop neutrino mass model based on an E6 Grand Unified Theory~\cite{Abdullah:2019ofw}. In this work, we will show that, in the MSSM with local $B-L$ gauge symmetry (B-LSSM)~\cite{FileviezPerez:2008sx,5,6}, without introducing explicit flavor mixing and requiring smuons are much heavier than selectrons, approximate values of the trilinear scalar terms $T_e$ in the soft supersymmetry breaking potential, slepton mass term $M_E$ and $\tan\beta$ can also account for the discrepancies. In addition, with respect to the MSSM, the effects of new parameters in the B-LSSM are also explored.

It is general believed that the SM is only the low energy approximation of a more fundamental,  unified theory. When $B-L$ symmetry~\cite{Pati:1974yy,Weinberg:1979sa,Davidson:1978pm,Mohapatra:1980qe,Marshak:1979fm,Wetterich:1981bx} is introduced, where $B$ represents the baryon number and $L$ represents the lepton number respectively, the corresponding heavy neutral vector boson can be considered as a possible remnant of unification~\cite{Buchmuller:1991ce}. The cosmological baryon asymmetry at temperatures much below the grand unified mass with spontaneously broken local $B-L$ symmetry are analyzed in Refs.~\cite{Masiero:1982fi,Mohapatra:1982xz}. In this work, we focus on the B-LSSM which can be obtained by extending the MSSM with local $B-L$ gauge symmetry. Compared with the MSSM, the gauge symmetry group of B-LSSM is extended to $SU(3)\otimes SU(2)_L\otimes U(1)_Y\otimes U(1)_{B-L}$. The invariance under the additional gauge group $U(1)_{B-L}$ imposes the R-parity conservation which is assumed in the MSSM to avoid proton decay. And R-parity conservation can be maintained if $U(1)_{B-L}$ symmetry is broken spontaneously~\cite{Das:2017flq}. $U(1)_{B-L}$ symmetry is broken by two additional Higgs singlets that carry $B-L$ charge, and the large Majorana masses for the right-handed neutrinos are generated by these Higgs fields. Combining with the Dirac mass term, three neutrinos obtain tiny masses by the see-saw mechanism, which can explain the tiny neutrino masses naturally ~\cite{Khalil:2006yi}. The model can also help to understand the origin of R-parity and its possible spontaneous violation in the supersymmetric models~\cite{Ashtekar:2007em,Barger:2008wn,Dulaney:2010dj}. Since the $B-L$ symmetry is radiatively broken at TeV scale, the model can implement the soft leptogenesis naturally~\cite{Babu:2009pi,Pelto:2010vq}. In addition, there are much more candidates for the dark matter (DM) in comparison to the MSSM: new neutralinos corresponding to the gauginos of $U(1)_{B-L}$ and additional Higgs singlets, as well as CP-even and -odd sneutrinos, the relic density and annihilations of these new DM candidates have been studied in Refs.~\cite{16,1616,DelleRose:2017ukx,DelleRose:2017uas}. Since both the additional Higgs singlets and right-handed (s)neutrinos release additional parameter space from the LEP, Tevatron and LHC constraints, the little hierarchy problem of the MSSM is also alleviated~\cite{search,77,88,9,99,10,11}.

The paper is organized as follows. In Sec.II, the B-LSSM and the contributions to $\bigtriangleup a_l^{NP}$ are discussed briefly. Then we explore the effects of $T_e$, $M_E$, $\tan\beta$ and new parameters in the B-LSSM on $\bigtriangleup a_{\mu,e}^{NP}$ by varying the values of them, in Sec.III. Conclusions are summarized in Sec.IV.

\section{B-LSSM and the contributions to $\bigtriangleup a_l^{NP}$\label{sec2}}
\begin{figure}
\setlength{\unitlength}{1mm}
\centering
\includegraphics[width=5in]{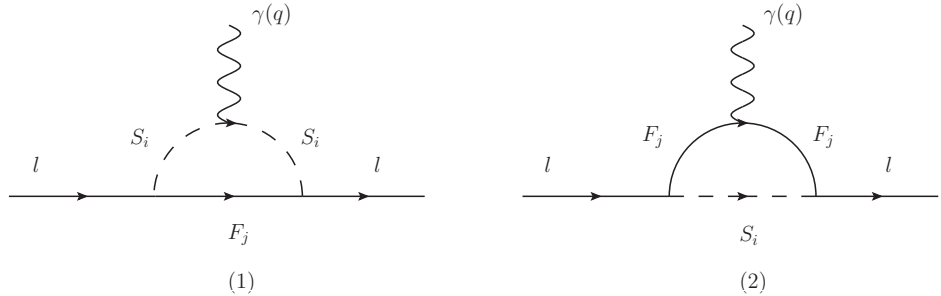}
\vspace{0cm}
\caption[]{Feynman diagrams contribute to the lepton MDM. (1) represents the contributions to $\bigtriangleup a_l^{NP}$ from charged scalars, while (2) represents the contributions from charged fermions}
\label{one loop Feynman diagram}
\end{figure}
\begin{figure}
\setlength{\unitlength}{1mm}
\centering
\includegraphics[width=6in]{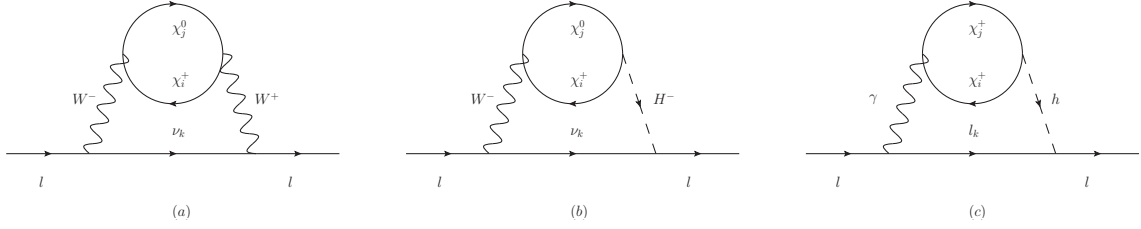}
\vspace{0cm}
\caption[]{The two-loop Barr-Zee type diagrams contribute to the lepton MDM, the corresponding contributions to $\bigtriangleup a_l^{NP}$ are obtained by attaching a photon to the internal particles in all possible ways.}
\label{two loop Feynman diagram}
\end{figure}
In the B-LSSM, the dominant contributions to lepton MDMs at the one-loop level come from the chargino-sneutrino loop (charginos, sneutrinos are loop particles) and the neutralino-slepton loop (neutralinos, sleptons are loop particles). Then the lepton MDM can be written as $a=a^n+a^c$, where $a^n$ denotes the lepton MDM results from the neutralino-slepton loop, and $a^c$ denotes the lepton MDM results from the chargino-sneutrino loop. In our previous work~\cite{Yang:2018guw}, we have discussed the muon MDM, and some two-loop Barr-Zee type diagrams are considered. The results show that the two-loop Barr-Zee type diagrams can make important corrections to the muon MDM. In this work, we consider the two-loop Barr-Zee type corrections, the corresponding one-loop and two-loop diagrams are depicted in Fig.~\ref{one loop Feynman diagram} and Fig.~\ref{two loop Feynman diagram} respectively. In the following analysis, we adopt the formulas in our previous work. In this sector, we present the dominant differences between the B-LSSM with the MSSM, and the new contributions to lepton MDMs in the B-LSSM are discussed.

In the B-LSSM, the chiral superfields and their quantum numbers are listed in Table. \ref{BLSSM}.
\begin{table*}
\begin{tabular*}{\textwidth}{@{\extracolsep{\fill}}cccc@{}}
\hline
superfields & Spin 0 & Spin $\frac{1}{2}$ & $U(1)_Y\bigotimes SU(2)_L\bigotimes SU(3)_C\bigotimes U(1)_{B-L}$\\
\hline
$\hat Q$  & $\tilde Q$ & $Q$   & $(\frac{1}{6},\bf2,\bf3,\frac{1}{6})$\\
$\hat D$  & $\tilde d^c$ & $d^c$   & $(\frac{1}{3},\bf1,\bar{\bf3},-\frac{1}{6})$\\
$\hat U$  & $\tilde u^c$ & $u^c$   & $(-\frac{2}{3},\bf1,\bar{\bf3},-\frac{1}{6})$\\
$\hat L$  & $\tilde L$ & $L$   & $(-\frac{1}{2},\bf2,\bf1,-\frac{1}{2})$\\
$\hat E$  & $\tilde e^c$ & $e^c$   & $(1,\bf1,\bf1,\frac{1}{2})$\\
$\hat \nu$  & $\tilde \nu^c$ & $\nu^c$   & $(0,\bf1,\bf1,\frac{1}{2})$\\
$\hat H_1$  & $H_1$ & $\tilde H_1$   & $(-\frac{1}{2},\bf2,\bf1,0)$\\
$\hat H_2$  & $H_2$ & $\tilde H_2$   & $(\frac{1}{2},\bf2,\bf1,0)$\\
$\hat \eta_1$  & $H_1$ & $\tilde \eta_1$   & $(0,\bf1,\bf1,-1)$\\
$\hat \eta_2$  & $H_2$ & $\tilde \eta_2$   & $(0,\bf1,\bf1,1)$\\
\hline
\end{tabular*}
\caption{Chiral superfields and their quantum numbers in the B-LSSM~\cite{OLeary:2011vlq}.}
\label{BLSSM}
\end{table*}
From the table we can see that two chiral singlet superfields $\hat{\eta}_{1}$, $\hat{\eta}_{2}$ and three generations of right-handed neutrinos are introduced in the B-LSSM, which allows for a spontaneously broken $U(1)_{B-L}$ without necessarily breaking R-parity. And the superpotential of the B-LSSM can be written as
\begin{eqnarray}
&&W=W^{MSSM}+Y_{\nu, ij}\hat{L_i}\hat{H_2}\hat{\nu}_j-\mu' \hat{\eta}_1 \hat{\eta}_2
+Y_{x, ij} \hat{\nu}_i \hat{\eta}_1 \hat{\nu}_j,\label{Eq4}
\end{eqnarray}
where $W^{MSSM}$ is the superpotential of the MSSM. There is a $\bigtriangleup L=2$ trilinear soft
breaking term $Y_{x, ij} \hat{\nu}_i \hat{\eta}_1 \hat{\nu}_j$ in the B-LSSM, which leads to a splitting between the real and imaginary parts of the sneutrino. As a result, there are twelve states in the sneutrino sector: six scalar sneutrinos and six pseudoscalar ones~\cite{Hirsch:1997vz,Grossman:1997is}. Eq. (\ref{Eq4}) shows that the right handed neutrinos obtain large Majorana masses since the expected size of the $u_{1,2}$ is $\sim10\;{\rm TeV}$, while the Dirac masses can be obtained by the terms $Y_{\nu, ij}\hat{L_i}\hat{H_2}\hat{\nu}_j$. Then three neutrinos obtain tiny masses naturally by the see-saw mechanism, and the neutrino Yukawa couplings do not have to be tiny to gain accord with neutrino mass limits. In addition, sneutrino masses are enlarged by the additional superpartners of the right-hand neutrinos in the B-LSSM, which plays a suppressive role to the contributions to lepton MDMs from the chargino-sneutrino loop, according to the decoupling theorem. Then the soft breaking terms of the B-LSSM are generally given as
\begin{eqnarray}
&&\mathcal{L}_{soft}=\mathcal{L}_{soft}^{MSSM}+\Big[-\frac{1}{2}(2M_{BB'}\tilde{\lambda}_{B'} \tilde{\lambda}_{B}+M_{B'}\tilde{\lambda}_{B'} \tilde{\lambda}_{B'})-
B_{\mu'}\tilde{\eta}_1 \tilde{\eta}_2+T_{\nu}^{ij} H_2 \tilde{\nu}_i^c \tilde{L}_j+\nonumber\\
&&\hspace{1.4cm}
T_x^{ij} \tilde{\eta}_1 \tilde{\nu}_i^c \tilde{\nu}_j^c+h.c.\Big]-m_{\tilde{\eta}_1}^2 |\tilde{\eta}_1|^2-
m_{\tilde{\eta}_2}^2 |\tilde{\eta}_2|^2,
\end{eqnarray}
where $\mathcal{L}_{soft}^{MSSM}$ is the soft breaking terms of the MSSM, $\tilde\lambda_{B}, \tilde\lambda_{B'}$ represent the gauginos of $U(1)_Y$, $U(1)_{(B-L)}$ correspondingly, and $M_{B'}$ is the $B'$ gaugino mass. Compared with the MSSM, there are three additional neutralinos in the B-LSSM, which can make contributions to lepton MDMs through the neutralino-slepton loop, and the two-loop Barr-Zee type diagrams shown in Fig.~\ref{two loop Feynman diagram}(a), (b). In addition, as the Higgs fields receive vacuum expectation values~\cite{Yang:2018utw}:
\begin{eqnarray}
&&H_1^1=\frac{1}{\sqrt2}(v_1+{\rm Re}H_1^1+i{\rm Im}H_1^1),
\qquad\; H_2^2=\frac{1}{\sqrt2}(v_2+{\rm Re}H_2^2+i{\rm Im}H_2^2),\nonumber\\
&&\tilde{\eta}_1=\frac{1}{\sqrt2}(u_1+{\rm Re}\tilde{\eta}_1+i{\rm Im}\tilde{\eta}_1),
\qquad\;\quad\;\tilde{\eta}_2=\frac{1}{\sqrt2}(u_2+i{\rm Re}\tilde{\eta}_2+i{\rm Im}\tilde{\eta}_2)\;,
\end{eqnarray}
the local gauge symmetry $SU(2)_L\otimes U(1)_Y\otimes U(1)_{B-L}$ breaks down to the electromagnetic symmetry $U(1)_{em}$. Conveniently, we can define $u^2=u_1^2+u_2^2,\; v^2=v_1^2+v_2^2$ and $\tan\beta'=\frac{u_2}{u_1}$ in analogy to the ratio of the MSSM VEVs ($\tan\beta=\frac{v_2}{v_1}$). $\tan\beta'$ appears in the mass matrix of slepton, which indicates that $\tan\beta'$ can affect the numerical results through the neutralino-slepton loop by affecting the slepton masses.

In the B-LSSM, there is a new gauge group $U(1)_{B-L}$, which introduces a new gauge coupling constant $g_{_B}$ and new gauge boson $Z'$. The updated experimental data~\cite{newZ} shows that, the new gauge boson mass $M_{Z'}\geq4.05\;{\rm TeV}$ at $95\%$ Confidence Level (CL). And an upper bound on the ratio between $M_{Z'}$ and $g_{_B}$ at $99\%$ CL is given in Refs.~\cite{20,21} as $M_{Z'}/g_B>6\;{\rm TeV}$. In addition, since there are two Abelian groups in the B-LSSM, and the invariance principle allows the Lagrangian to include a mixing term between the strength tensors of gauge fields corresponding to the two Abelian groups, a new effect arises in the B-LSSM: the gauge kinetic mixing. Then the form of covariant derivatives can be redefined as
\begin{eqnarray}
&&D_\mu=\partial_\mu-i\left(\begin{array}{cc}Y,&B-L\end{array}\right)
\left(\begin{array}{cc}g_{_Y},&g_{_{YB}}^\prime\\g_{_{BY}}^\prime,&g_{_{B-L}}\end{array}\right)
\left(\begin{array}{c}A_{_\mu}^{\prime Y} \\ A_{_\mu}^{\prime BL}\end{array}\right)\;.
\label{gauge1}
\end{eqnarray}
As long as the two Abelian gauge groups are unbroken, the basis can be changed as:
\begin{eqnarray}
&&D_\mu=\partial_\mu-i\left(\begin{array}{cc}Y,&B-L\end{array}\right)
\left(\begin{array}{cc}g_{_Y},&g_{_{YB}}^\prime\\g_{_{BY}}^\prime,&g_{_{B-L}}\end{array}\right)R^TR
\left(\begin{array}{c}A_{_\mu}^{\prime Y} \\ A_{_\mu}^{\prime BL}\end{array}\right)\nonumber\\
&&\quad\;\;=\partial_\mu-i\left(\begin{array}{cc}Y,&B-L\end{array}\right)
\left(\begin{array}{cc}g_{_1},&g_{_{YB}}\\0,&g_{_{B}}\end{array}\right)
\left(\begin{array}{c}A_{_\mu}^{Y} \\ A_{_\mu}^{BL}\end{array}\right)
\label{gauge2}
\end{eqnarray}
where $R$ is a $2\times2$ orthogonal matrix. As a result, gauge mixing is introduced in various kinetic terms of Lagrangian by the new definition of covariant derivatives. And interesting consequences of the gauge kinetic mixing arise in various sectors of the model. Firstly, new gauge coupling constant $g_{_{YB}}$ is introduced, and new gauge boson $Z'$ mixes with the $Z$ boson in the MSSM at the tree level. Correspondingly, new gaugino $\tilde\lambda_{B'}$ also mixes with bino at the tree level, the mixing mass term $M_{BB'}$ is introduced. Then the gauge kinetic mixing leads to the mixing between $H_1^1,\;H_2^2,\;\tilde{\eta}_1,\;\tilde{\eta}_2$ at the tree level, and $\tilde\lambda_{B'}$ mixes with the two higgsinos in the MSSM, which means that the new gauge coupling constant $g_{_{YB}}$ can affect the numerical results through the neutralino-slepton loop. Meanwhile, additional D-terms contribute to the mass matrices of sleptons. On the basis $(\tilde L, \tilde e^c)$, the slepton mass matrix is given by
\begin{eqnarray}
&&m_{\tilde e}^2=
\left(\begin{array}{cc}m_{eL}^2,&\frac{1}{\sqrt2}(v_1 T_e^\dagger-v_2\mu Y_e^\dagger)\\\frac{1}{\sqrt2}(v_1 T_e-v_2\mu^* Y_e),&m_{eR}^2\end{array}\right),\label{ME1}
\end{eqnarray}
\begin{eqnarray}
&&m_{eL}^2=\frac{1}{8}\Big[2g_{_B}(g_{_B}+g_{_{YB}})(u_1^2-u_2^2)+(g_1^2-g_2^2+g_{_{YB}}^2+2g_{_B}g_{_{YB}})(v_1^2-
v_2^2)\Big]\nonumber\\
&&\qquad\;\quad\;+m_{\tilde L}^2+\frac{v_1^2}{2}Y_e^\dagger Y_e,\nonumber\\
&&m_{eR}^2=\frac{1}{24}\Big[2g_{_B}(g_{_B}+2g_{_{YB}})(u_2^2-u_1^2)+2(g_1^2+g_{_{YB}}^2+2g_{_B}g_{_{YB}})(v_2^2-
v_1^2)\Big]\nonumber\\
&&\qquad\;\quad\;+m_{\tilde e}^2+\frac{v_1^2}{2}Y_e^\dagger Y_e.\label{ME2}
\end{eqnarray}
It can be noted that $\tan\beta'$ and new gauge coupling constants $g_{_B}$, $g_{_{YB}}$ in the B-LSSM can affect numerical results by affecting the slepton masses.


\section{Numerical analyses\label{sec3}}

The numerical results of $\bigtriangleup a_\mu^{NP}$ and $\bigtriangleup a_e^{NP}$ are displayed in this section. The relevant SM input parameters are chosen as $m_W=80.385\;{\rm GeV},\;m_Z=90.1876\;{\rm GeV},\;m_e=5.11\times10^{-4}\;{\rm GeV},\;m_\mu=0.105\;{\rm GeV},\;\alpha_{em}(m_Z)=1/128.9$. Since the tiny neutrino masses affect the numerical analysis negligibly, we take $Y_\nu=Y_x=0$ approximately.

Since the contribution of heavy $Z'$ boson is highly suppressed, we take $M_{Z'}=4.2\;{\rm TeV}$ in the following analysis. In our previous work~\cite{JLYang:2018}, the rare processes $\bar B\rightarrow X_s\gamma$ and $B_s^0\rightarrow \mu^+\mu^-$ are discussed in detail, and we take the charged Higgs boson mass $M_{H^\pm}=1.5\;{\rm TeV}$ to satisfy the experimental data on these processes. In addition, in order to satisfy the constraints from the experiments~\cite{PDG}, for those parameters in higgsino, gaugino and sneutrino sectors, we appropriately fix $M_1=\frac{1}{2}M_2=\frac{1}{2}\mu=0.3\;{\rm TeV}$, $m_\nu=diag(1,1,1)\;{\rm TeV}$, $T_x=T_\nu=0.1\;{\rm TeV}$, for simplicity, where $m_\nu$ is the right-handed sneutrino soft mass matrix. All of the parameters fixed above affect the following numerical analysis negligibly. When the leading-log radiative corrections from stop and top particles are included~\cite{HiggsC1,HiggsC2,HiggsC3}, right SM-like Higgs boson mass can be obtained with appropriate parameters in squark sector, which is irrelevant with the theoretical predictions of lepton MDMs. The nature of DM candidate, the sneutrino in the B-LSSM, has been studied in Ref.~\cite{DelleRose:2017uas}, the results show that the sneutrino masses in our chosen parameter space can obtain right DM abundance. Furthermore, we take soft breaking slepton mass matrix $m_{\tilde L, \tilde e}=diag(M_E, M_E, M_E)$ and the trilinear coupling matrix $T_e=diag(A_L, A_L, A_L)$, where $T_e=A_L\times Y_e$ is not employed in our definition. In order to conveniently discuss the discrepancies between $\bigtriangleup a_\mu^{NP}$ and $\bigtriangleup a_e^{NP}$, we define
\begin{eqnarray}
&&R_\mu=\frac{\bigtriangleup a_\mu^{NP}\times 10^9-2.74}{0.73},\label{dRmu}\\
&&R_e=\frac{\bigtriangleup a_e^{NP}\times 10^{13}+8.8}{3.6}.\label{dRe}
\end{eqnarray}
It is obvious that $R_{\mu,e}$ denote the standard deviations between the B-LSSM predictions and experiments. And $R_{\mu,e}=0$ indicates that the theoretical predictions on $a_{\mu,e}$ are at the corresponding experimental central values, when the NP contributions are considered.

\begin{figure}
\setlength{\unitlength}{1mm}
\centering
\includegraphics[width=4in]{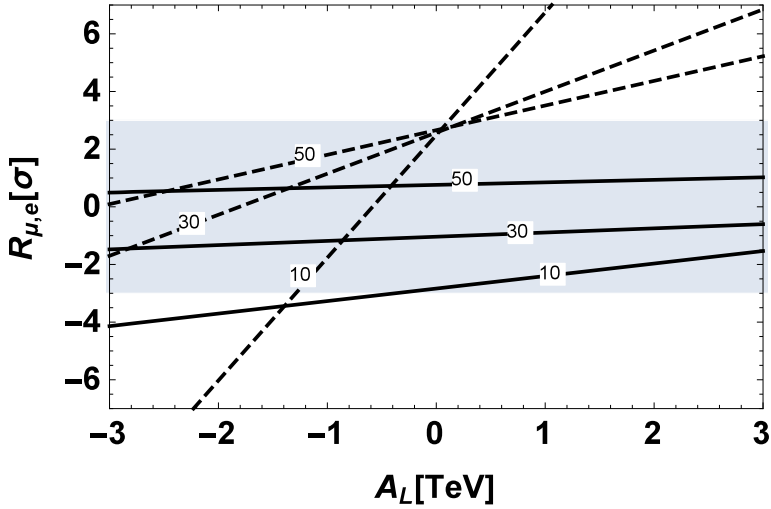}
\vspace{0cm}
\caption[]{Taking $M_{B'}=M_{BB'}=0.6\;{\rm TeV}$, $\mu'=0.8\;{\rm TeV}$, $g_{_B}=0.4$, $g_{_{YB}}=-0.4$, $\tan\beta'=1.15$, $M_E=1.5\;{\rm TeV}$, $R_\mu$ (solid lines) and $R_e$ (dashed lines) versus $A_L$ for $\tan\beta=10, 30, 50$ are plotted, where the gray area denotes the experimental $3\sigma$ interval.}
\label{Rl-Ae}
\end{figure}

Then taking $M_{B'}=M_{BB'}=0.6\;{\rm TeV}$, $\mu'=0.8\;{\rm TeV}$, $g_{_B}=0.4$, $g_{_{YB}}=-0.4$, $\tan\beta'=1.15$, $M_E=1.5\;{\rm TeV}$, we present $R_\mu$ (solid lines) and $R_e$ (dashed lines) versus $A_L$ in Fig.~\ref{Rl-Ae} for $\tan\beta=10, 30, 50$, where the gray area denotes the experimental $3\sigma$ interval. In the plotting, we adopt $R_{\mu,e}$ defined in Eq.~(\ref{dRmu}), (\ref{dRe}) respectively as $y$-axis, without changing anything. And Eq.~(\ref{dRmu}), (\ref{dRe}) show that $R_\mu\simeq-3.7$ and $R_e\simeq2.4$ when $\bigtriangleup a_{\mu,e}^{NP}=0$. Combining Eq.~(\ref{ME1}), (\ref{ME2}) and the concrete expressions of lepton MDM at the one-loop level in our previous work~\cite{Yang:2018guw}, we can see that, if we do not count the suppressive factor $m_l^2$, the dominant contribution from the neutralino-slepton loop $a^n$ is proportional to $(v A_L/\tan\beta-\sqrt{2}\mu\tan\beta m_l)/\Big(m_l\sqrt{M_{LR}^2+(v A_L/\tan\beta-\sqrt{2}\mu\tan\beta m_l)^2}\Big)$ approximately, where $M_{LR}=(m_{eL}^{l2}-m_{eR}^{l2})/\sqrt2$. And the dominant contribution from the chargino-sneutrino loop $a^c$ is proportional to $\tan\beta$ approximately. Hence, the contributions from $a^n$ are negative when $A_L$ is negative, and the sign of $a_n$ can be changed when $vA_L/\tan\beta>\sqrt{2}\mu\tan\beta m_l$. For $\bigtriangleup a_{e}^{NP}$, the dominant contributions come from $a^n$, hence the NP contributions to $\bigtriangleup a_{e}^{NP}$ are negative when $vA_L/\tan\beta<\sqrt{2}\mu\tan\beta m_l$, and positive when $vA_L/\tan\beta>\sqrt{2}\mu\tan\beta m_l$, approximately. As we can see from the picture, the NP contributions to $\bigtriangleup a_{e}^{NP}$ are negative when $A_L\lesssim-0.02\;{\rm TeV}$ for $\tan\beta=10$, $A_L\lesssim-0.1\;{\rm TeV}$ for $\tan\beta=30$, $A_L\lesssim-0.3\;{\rm TeV}$ for $\tan\beta=50$, and the NP contributions to $\bigtriangleup a_{e}^{NP}$ are positive when the values of $A_L$ are larger than these values correspondingly. And it is obvious that the maximum value of $A_L$ increases with the increasing of $\tan\beta$ when the NP contributions to $\bigtriangleup a_{e}^{NP}$ are negative, which results from that $a^n$ is suppressed by large $\tan\beta$, while $a^c$ is enhanced by large $\tan\beta$, and the signs of $a^n$, $a^c$ are opposite in this case.

When $A_L=-3\;{\rm TeV}$, $\tan\beta=10$, if we do not count the suppressive factor $m_l^2$, the dominant contributions to $\bigtriangleup a_{\mu,e}^{NP}$ come from the neutralino sector, which are negative and have a enhancing factor $1/m_{\mu,e}$, hence the contributions to $\bigtriangleup a_{e}^{NP}$ is larger than $\bigtriangleup a_{\mu}^{NP}$. As we can see from the picture, $\bigtriangleup a_{\mu}^{NP}$ receives quite small and negative contributions when $A_L=-3 {\rm TeV}$, $\tan\beta=10$, while $\bigtriangleup a_{e}^{NP}$ receives quite large and negative contributions. In addition, when $A_L=-3 {\rm TeV}$, $\tan\beta=30, 50$, the contributions from $a^n$ have a suppressive factor $1/\tan\beta$, while the contributions from $a^c$ are enlarged by large $\tan\beta$. For $\bigtriangleup a_{e}^{NP}$, $a^n$ is enhanced vastly by $1/m_e$, hence even $a^n$ is suppressed by $1/\tan\beta$ and $a^c$ is enhanced by $\tan\beta$, the contributions from $a^n$ are still larger than $a^c$. As we can see from the picture, $\bigtriangleup a_{e}^{NP}$ is negative and decreases with the increasing of $\tan\beta$ when $A_L=-3\;{\rm TeV}$. But for $\bigtriangleup a_{\mu}^{NP}$, the enhancing factor of $a^n$ is $1/m_\mu<1/m_e$, hence the contributions from $a^c$ are larger than $a^n$ when $\tan\beta=30, 50$, and $\bigtriangleup a_{\mu}^{NP}$ receives positive contributions in this case. And $R_\mu\approx R_e$ when $\tan\beta=30, 50$ does not indicate $\bigtriangleup a_{\mu}^{NP}\approx\bigtriangleup a_{e}^{NP}$, if we do not count the suppressive factor $m_l^2$, the contributions to $\bigtriangleup a_{e}^{NP}$ are negative, while the contributions to $\bigtriangleup a_{\mu}^{NP}$ are positive.

If we limit the NP corrections to $\bigtriangleup a_{\mu,e}^{NP}$ in $3\sigma$ interval, the experimental results prefer $A_L\lesssim0.4\;{\rm TeV}$ for $\tan\beta=30,50$, and $-0.4\lesssim A_L\lesssim0.1\;{\rm TeV}$ for $\tan\beta=10$. It can be noted that, the allowed region of $A_L$ for $\tan\beta=10$ is limited strictly in our chosen parameter space. According to Ref.~\cite{Moroi:1995yh}, the contributions to $\bigtriangleup a_{\mu}^{NP}$ can be enhanced by large $\mu$. However, the allowed region of $A_L$ for $\tan\beta=10$ can be enlarged when $\mu\lesssim-20\;{\rm TeV}$ (the additional minus sign comes from the different definition of $\mu$ in Ref.~\cite{Moroi:1995yh}), which is not the region of $\mu$ we are interested in. And $\mu$ appears in the expression of $a^n$ as $\mu\times m_l$, the effect of $\mu$ to $\bigtriangleup a_{e}^{NP}$ is highly suppressed by small $m_e$, hence we do not discuss the effect of $\mu$ in the following analysis. In addition, it can be noted that $A_L$ affects the numerical results less obviously with the increasing of $\tan\beta$. Because $A_L$ affects the numerical results mainly by affecting the contributions of $a^n$, and $A_L$ appears in the expression as $A_L/\tan\beta$, which indicates that the effect of $A_L$ is suppressed by large $\tan\beta$.

\begin{figure}
\setlength{\unitlength}{1mm}
\centering
\includegraphics[width=4in]{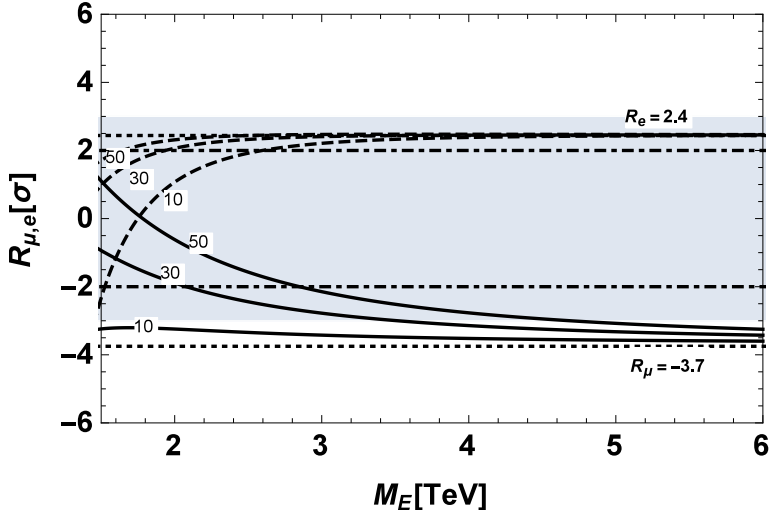}
\vspace{0cm}
\caption[]{Taking $M_{B'}=M_{BB'}=0.6\;{\rm TeV}$, $\mu'=0.8\;{\rm TeV}$, $g_{_B}=0.4$, $g_{_{YB}}=-0.4$, $\tan\beta'=1.15$, $A_L=-1\;{\rm TeV}$, $R_\mu$ (solid lines) and $R_e$ (dashed lines) versus $M_E$ for $\tan\beta=10, 30, 50$ are displayed, where the gray area denotes the experimental $3\sigma$ interval, the dotdashed line denote the experimental $2\sigma$ bounds, and the dotted lines denote the corresponding decoupling limits for $R_\mu$, $R_e$.}
\label{Rl-ME}
\end{figure}

Assuming $A_L=-1\;{\rm TeV}$, $R_\mu$ (solid lines) and $R_e$ (dashed lines) versus $M_E$ are plotted in Fig.~\ref{Rl-ME} for $\tan\beta=10, 30, 50$, where the gray area denotes the experimental $3\sigma$ interval, the dotdashed lines denote the experimental $2\sigma$ bounds, the dotted lines denote the corresponding decoupling limits for $R_\mu$ and $R_e$.
It can be noted in the picture that, with the increasing of $M_E$, the theoretical predictions on $R_\mu$ and $R_e$ decouple to the corresponding SM predictions, which coincides with the decoupling theorem. And in our chosen parameter space, the region of $M_E$ is excluded by $R_\mu$ for $\tan\beta=10$, if we limit the NP corrections to $\bigtriangleup a_{\mu}^{NP}$ in $3\sigma$ interval. In addition, if we limit the NP corrections to $\bigtriangleup a_{\mu,e}^{NP}$ in $2\sigma$ interval, the numerical results show that, $M_E$ is limited in the region $M_E\lesssim2\;{\rm TeV}$ for $\tan\beta=30$ and $M_E\lesssim1.7\;{\rm TeV}$ for $\tan\beta=50$.

\begin{figure}
\setlength{\unitlength}{1mm}
\centering
\includegraphics[width=3.1in]{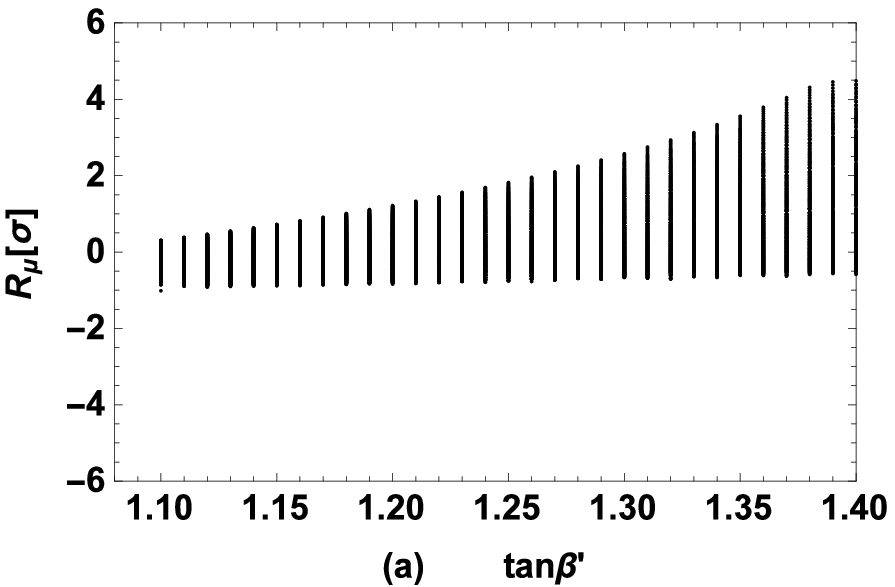}
\vspace{0.5cm}
\includegraphics[width=3.1in]{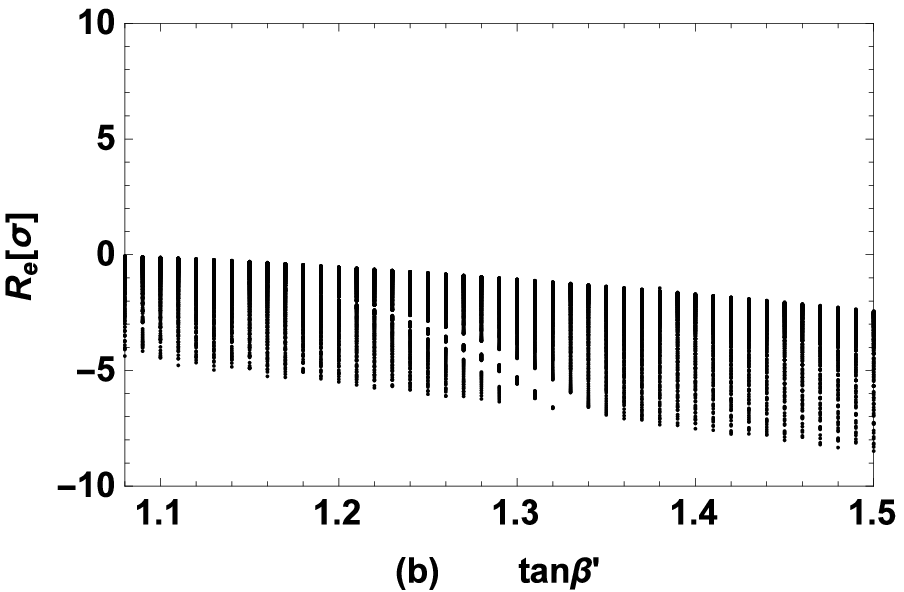}
\vspace{0cm}
\caption[]{Taking $\tan\beta=30$, $M_E=1.2\;{\rm TeV}$, $M_{B'}=M_{BB'}=0.6\;{\rm TeV}$, $\mu'=0.8\;{\rm TeV}$, and scanning $\tan\beta'$ in the range $(1.02\sim1.5)$, $g_{_B}$ in the range $(0.1\sim0.7)$, $g_{_{YB}}$ in the range $(-0.7\sim-0.1)$, then $R_\mu$ (a) and $R_e$ (b) versus $\tan\beta'$ are plotted.}
\label{scan1}
\end{figure}

Compared with the MSSM, there are some new parameters in the B-LSSM, we take $\tan\beta=30$, $M_E=1.2\;{\rm TeV}$, $M_{B'}=M_{BB'}=0.6\;{\rm TeV}$, $\mu'=0.8\;{\rm TeV}$, and scan the parameter space shown in Table~\ref{tab1}.
\begin{table*}
\begin{tabular*}{\textwidth}{@{\extracolsep{\fill}}llll@{}}
\hline
parameters & min & max & step\\
\hline
$\tan\beta'$  & 1.02 & 1.5  & 0.01\\
$g_{_B}$      & 0.1  &  0.7 & 0.02\\
$g_{_{YB}}$   & -0.7 & -0.1 & 0.02\\
\hline
\end{tabular*}
\caption{Taking $\tan\beta=30$, $M_E=1.2\;{\rm TeV}$, $M_{B'}=M_{BB'}=0.6\;{\rm TeV}$, $\mu'=0.8\;{\rm TeV}$, the scanning parameters for Fig. \ref{scan1}.}
\label{tab1}
\end{table*}
In the scanning, we keep the slepton masses $m_{L_i}>500\;{\rm GeV}(i=1,\cdot\cdot\cdot,6)$, the Higgs boson mass in experimental $3\sigma$ interval, to avoid the range ruled out by the experiments\cite{PDG}. Then we plot $R_\mu$ versus $\tan\beta'$ in Fig.~\ref{scan1} (a), $R_e$ versus $\tan\beta'$ in Fig.~\ref{scan1} (b).
The picture shows that, $R_\mu$ increases with the increasing of $\tan\beta'$, while $R_e$ decreases with the increasing of $\tan\beta'$, which indicates that $\tan\beta'$, $g_{_B}$, $g_{_{YB}}$ can affect the numerical results, and the effects of them are comparable. Due to our definition of $R_{\mu,e}$, both $\bigtriangleup a_{\mu}^{NP}$ and $\bigtriangleup a_{e}^{NP}$ increase with the increasing of $\tan\beta'$. Eq.~(\ref{ME2}) shows that the lepton masses decrease with the increasing of $\tan\beta'$ when $|g_{_{YB}}|<g_{_B}<2|g_{_{YB}}|$, which indicates that the theoretical predictions on $\bigtriangleup a_{\mu,e}^{NP}$ can be enhanced by large $\tan\beta'$ in this case. In addition, it can be noted that the NP contributions to the muon MDM are positive, while the NP contributions to the electron MDM are negative, in our chosen parameter space. It results from that, when $\tan\beta=30$, the contributions from $a^n$ to $\bigtriangleup a_{l}^{NP}$ are proportional to $\frac{1}{m_l \tan\beta}$ approximately, while the contributions from $a^c$ are proportional to $\tan\beta$. And when $A_L<0\;{\rm TeV}$, $a^n$ is negative, $a^c$ is positive. For $\bigtriangleup a_{e}^{NP}$, although $a^n$ is suppressed by $1/\tan\beta$, and $a^c$ is enhanced by $\tan\beta$, when $\tan\beta=30$, but the enhancing factor $1/m_e$ is large enough to have $|a^n|>a^c$, hence the contributions to $\bigtriangleup a_{e}^{NP}$ are negative. But for $\bigtriangleup a_{\mu}^{NP}$, the enhancing factor $1/m_\mu$ is not large enough to have $|a^n|>a^c$ in this case, and as a result, the contributions to $\bigtriangleup a_{\mu}^{NP}$ are positive.

In the B-LSSM, there are three additional mass terms in the neutralino sector. In order to see how $M_{BB'}$, $M_{B'}$ and $\mu'$ affect the theoretical predictions on $\bigtriangleup a_{\mu,e}^{NP}$, we take $\tan\beta'=1.15$, $g_{_B}=0.4$, $g_{_{YB}}=-0.4$, and scan the parameter space shown in Table~\ref{tab2}.
\begin{table*}
\begin{tabular*}{\textwidth}{@{\extracolsep{\fill}}llll@{}}
\hline
parameters & min & max & step\\
\hline
$M_{BB'}$[TeV]  & 0   & 3  & 0.1\\
$M_{B'}$[TeV]   & 0   & 3  & 0.1\\
$\mu'$[TeV]     & 0.1 & 3  & 0.1\\
\hline
\end{tabular*}
\caption{Taking $\tan\beta'=1.15$, $g_{_B}=0.4$, $g_{_{YB}}=-0.4$, the scanning parameters for Fig. \ref{scan2}.}
\label{tab2}
\end{table*}
\begin{figure}
\setlength{\unitlength}{1mm}
\centering
\includegraphics[width=3.1in]{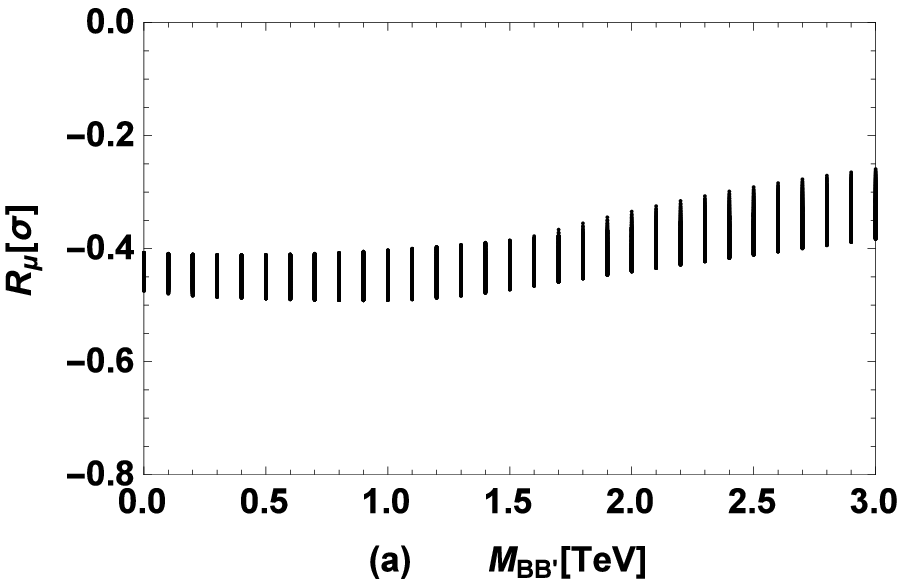}
\vspace{0.5cm}
\includegraphics[width=3.1in]{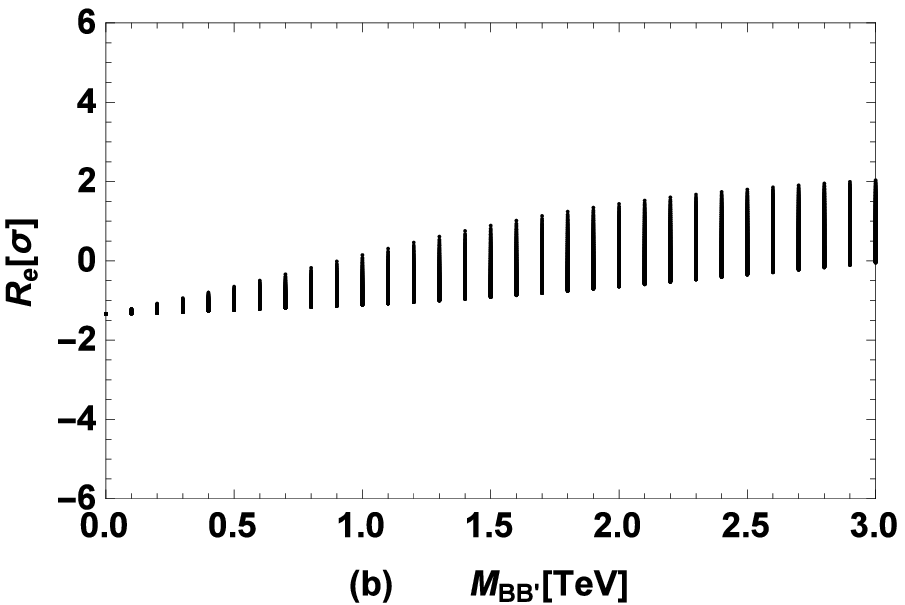}
\vspace{0cm}
\caption[]{Taking $\tan\beta'=1.15$, $g_{_B}=0.4$, $g_{_{YB}}=-0.4$, and scanning $M_{BB'}$, $M_{B'}$ in the range $0\sim3$ TeV, $\mu'$ in the range $0.1\sim3$ TeV, then $R_\mu$ (a) and $R_e$ (b) versus $M_{BB'}$ are plotted.}
\label{scan2}
\end{figure}
It can be noted in the table that, we take the minimum values of $M_{BB'}$ and $M_{B'}$ equal to $0\;{\rm TeV}$, because the gaugino masses still can be large enough to satisfy the experimental upper bounds on gaugino masses even if the values of $M_{BB'}$ and $M_{B'}$ are very small. Then we plot $R_\mu$ and $R_e$ versus $M_{BB'}$ in Fig. \ref{scan2} (a), (b) respectively. In the scanning, we keep the gaugino masses $>100{\rm GeV}$, to avoid the range ruled out by the experiments. From the picture we can see that, in our chosen parameter space, both $R_\mu$ and $R_e$ are in the experimental $2\sigma$ interval with the changing of new parameters $M_{BB'}$, $M_{B'}$ and $\mu'$. In addition, $M_{B'}$ and $\mu'$ affect the numerical results more obviously with larger $M_{BB'}$. Because $M_{BB'}$ is the mixing term between $\tilde \lambda_B$ and $\tilde \lambda_{B'}$, the mixing between $\tilde \lambda_B$ and $\tilde \lambda_{B'}$ is stronger with larger $M_{BB'}$, which leads that $M_{B'}$ can affect the numerical results more obviously. As a result, three additional mass terms in the neutralino sector of B-LSSM can affect the theoretical predictions on $R_\mu$ and $R_e$.

\section{Summary\label{sec4}}

In the frame work of B-LSSM, we focus on the muon and electron discrepancies, which results from a recent improved determination of the fine structure constant. And in the calculation, some two-loop Barr-Zee type diagrams are considered. Without introducing explicit flavor mixing and requiring smuons are much heavier than selectrons, we find that appropriate values of the trilinear scalar term $T_e$ in the soft supersymmetry breaking potential, slepton mass term $M_E$ and $\tan\beta$ can also account for the discrepancies. Considering the constraints from updated experimental data, the numerical results show that, if we limit the NP corrections to $\bigtriangleup a_{\mu,e}^{NP}$ in $2\sigma$ interval, the experimental results on $a_\mu$ and $a_e$ favor minus $T_e$, small $M_E$ ($M_E\lesssim2\;{\rm TeV}$) and large $\tan\beta$, in our chosen parameter space. In addition, there are new parameters $\tan\beta'$, $g_{_B}$, $g_{_{YB}}$, $M_{BB'}$, $M_{B'}$ and $\mu'$ in the B-LSSM with respect to the MSSM, all of them can affect the theoretical predictions on $\bigtriangleup a_{\mu,e}^{NP}$ through the neutralino-slepton loop, and $M_{BB'}$, $M_{B'}$, $\mu'$ can also make contributions to lepton MDMs through the considered two-loop Barr-Zee type diagrams.

\begin{acknowledgments}

The work has been supported by the National Natural Science Foundation of China (NNSFC) with Grants No. 11535002, and No. 11705045, the youth top-notch talent support program of the Hebei Province, Hebei Key Lab of Optic-Eletronic Information and Materials, and the Midwest Universities Comprehensive Strength Promotion project.
\end{acknowledgments}

\end{document}